# Enhanced THz emission and exciton transfer in monolayer MoS$_2$/GaAs heterostructures

C. Abinash Bhuyan*,[1,2], Anil K. Chaudhary*,[1], Kishore K. Madapu[3], P. Naveen Kumar[1], Sandip Dhara[3]

[1]Advanced Centre for Research in High Energy Materials, School of Physics, University of Hyderabad, Telangana, 500046, India

[2]Department of Physics, San Diego State University, San Diego, CA 92182, United States

[3]Condensed Matter Physics Division, Indira Gandhi Centre for Atomic Research, A CI of Homi Bhabha National Institute, Kalpakkam 603 102, India

*Email: cbhuyan@sdsu.edu, akcsp@uohyd.ernet.in, akcphys@gmail.com.

## Abstract

For designing an efficient terahertz (THz) emitter, the defect density of the semiconductors is smartly increased to reduce carrier lifetime, which subsequently lowers the overall power output of the semiconductor. To overcome this fundamental trade-off, this study presents a novel approach, by integrating a direct band gap 2D semiconductor such as monolayer MoS$_2$ (1L-MoS$_2$) with a well-known THz emitter, low-temperature-grown gallium arsenide (GaAs). The fabricated hybrid 2D/3D vertical van der Waals heterostructure showed a 15% higher THz emission compared to bare GaAs due to phase-coherent addition of second-order non-linear susceptibility, and overall enhancement in the electric field of laser. The photoluminescence (PL) enhancement factor of 2.38 in heterostructures at GaAs emission energy of 1.42 eV. However, the substantial quenching of PL emission for 1L-MoS$_2$ at the energy of 1.84 eV, is attributed to the Dexter-type exciton transfer mechanism at type-I band alignment. THz time-domain spectroscopy reveals a significant increase in optoelectronic properties, such as optical conductivity becoming doubled, and a 50% reduction in absorption coefficients. The study introduces a new route for fabricating large-area and compact mixed-dimensional van der Waals heterostructures, which can be used to enhance the efficiency of conventional semiconductor technologies and THz-based optoelectronic devices.

# 1. Introduction

Terahertz (THz) technologies fill the gap between microwave and optical frequencies, expanding the field of both photonics and electronics for advanced applications like wireless connectivity, artificial intelligence, imaging, and spectroscopy [1-3]. However, the development of efficient THz emitters has been hindered by the lack of suitable materials and technologies capable of generating and manipulating THz waves effectively [4, 5]. Currently, various devices such as free-electron lasers, quantum cascade lasers, photoconductive antennas (PCAs), and nonlinear organic and semiconductor crystals are extensively used to generate THz radiation [6, 7]. THz generation is mostly carried out using two primary methods: utilizing PCAs and bare semiconductor surfaces (BSS) [8, 9]. BSS is a more fundamental requirement because the addition of electrodes to it is used in fabricating PCAs to enhance its efficiency further. Moreover, BSS is advantageous as the entire surface contributes to THz generation and can withstand higher laser power compared to any fabricated antenna. Notably, BSS, utilizing common THz-active semiconductors such as GaAs, GaSb, InAs, InSb, and InP, is frequently employed for THz generation [10-12]. In a lower carrier concentration regime, electrons in GaAs behave as localized due to Coulomb interactions [13]. When defects are introduced to increase the carrier concentrations in GaAs, carrier lifetime decreases, and subsequently, the output emitted power also decreases further. Therefore, there is a demand for an increase in THz emissions in GaAs by adopting an innovative methodology, which is not only a fundamental interest but also a technological challenge by forming a heterojunction with a novel material.

In the literature on THz emitters, the power of emitted THz radiation in nonlinear optical (NLO) materials is higher for those exhibiting exciting optoelectronic properties comparable to their electronic counterparts [14, 15]. The existing technologies rely on bulk form of NLO materials and they demand multiple optical confinement methods for their photonics device applications. Unfortunately, conventional bulk forms of semiconductors have low NLO susceptibility [14, 16]. In contrast, recently emerged 2D materials exhibit remarkable optical, excitonic, and large-optical non-linearity [14, 15, 17, 18]. Among the 2D transition metal dichalcogenides (TMDs), monolayer $MoS_2$ (1L-$MoS_2$) is a well-known direct band gap ultra-thin semiconductor [19] which shows various promising electronic and optoelectronic properties [20-23]. The van der Waals nature of the surface and mechanical flexibility, make them suitable for integration into any existing technologies [24, 25]. The broad definition of a van der Waals heterostructure encompasses passivated, dangling bond-free 2D materials that engage in non-covalent interactions and interface with bulk inorganic semiconductors [26]. In particular, it was reported that 1L-MoS$_2$ shows an ultra-strong non-linear process due to the trigonal warping effect, which makes them suitable for THz emission [23]. Under lower carrier concentrations, electrons behave localized in GaAs or Si, whereas electrons in 1L-MoS$_2$ behave as 2D electron gas [13]. Similarly, a THz active semiconductor such as GaAs is widely used for emission applications. GaAs is capable of generating THz radiation efficiently, with specific mechanisms such as quantum interference and surface nonlinearity enhancing its performance [27, 28]. It is a direct band gap semiconductor that produces THz by optical rectification process and photo Dember effect [29, 30]. In particular, low-temperature GaAs show excellent electronic properties, high electron mobility, and good thermal stability. Particularly, the higher electron mobility of GaAs contributes to fast photoconductive rise time [31]. GaAs have superior dark resistivity and breakdown field [32]. Also, low-temperature-grown GaAs interdigitated photoconductive antennas are used to generate THz pulses [33]. It has been extensively employed in the development of optoelectronic devices, high-frequency electronics, and photovoltaic applications [30, 34]. Therefore, fabricating the van der Waals heterostructures with a THz active semiconductor is a novel approach for THz generation. Sometimes, heterostructures offer synergistic properties and new functionalities that are not exhibited by any constituted materials [35]. Also, current all-2D van der Waals heterostructures face performance limitations and higher fabrication costs compared to a 2D material integration with established semiconductor technologies [26]. The literature also lacks studies on the large-area integration of 2D TMDCs with GaAs semiconductors [26], particularly for applications in THz emission.

In this report, THz generation in GaAs is enhanced by 15% with the integration of a 2D electron gas material. For large-area integration, the mixed-dimensional van der Waals heterostructures of 1L-MoS$_2$/GaAs were fabricated using the transfer of 1L-MoS$_2$ film onto GaAs substrates, and the successful transfer of 1L-MoS$_2$ was confirmed by Raman spectroscopy. Photoluminescence (PL) analysis establishes the type-I band alignment and Dexter-type exciton transfer from the 1L-MoS$_2$ to GaAs. To investigate the photo-physics of heterostructures under femtosecond laser pulses, terahertz time-domain spectroscopic (THz-TDS) measurements were conducted. Enhanced THz generation was achieved through an amplified optical rectification process, which was systematically studied under pulsed laser excitation.

## 2. Results and Discussion

Unlike mechanically exfoliated flakes, large-area 2D films enable scalable manufacturing and compatibility with industry-standard semiconductor processes [26]. The large-area 1L-MoS$_2$ film was successfully grown using the atmospheric pressure CVD method. The similar CVD-grown film used in this study was also used for various optical studies [36-38]. The large-area MoS$_2$ films were obtained by optimizing the precursor amount, growth temperature, growth time, and carrier gas flow rate of the CVD system [39]. The schematic of the CVD setup is provided in Figure S1. The details of the CVD synthesis protocols, and optimized parameters are explicitly provided in the experimental method. The microscopic image of large-area 1L-MoS$_2$ film on SiO$_2$/Si substrate is shown in Figure1a. To show the optical contrast between the SiO$_2$ substrate and 1L-MoS$_2$, the optical image depicts both areas. Further, field-emission scanning electron microscopy (FESEM) measurements were carried out, and found that the surface is uniform. Similar to the optical image, the FESEM image was also collected, encompassing both SiO$_2$ and 1L-MoS$_2$ surfaces. The FESEM image reveals the uniform growth of MoS$_2$ film with few grain boundaries (Figure 1b). It is obvious that a large-area 1L-MoS$_2$ film contains grain boundaries because the constituent 1L-MoS$_2$ flakes are merged to form the large area. To further elucidate the composition, phase, and number of layers grown, Raman measurements were performed. Detailed measurement protocols are provided in the experimental section. In the back-scattering configuration, 1L-MoS$_2$ shows two prominent Raman modes of $E^1_{2g}$ and $A_{1g}$ phonons, which correspond to out-of-phase in-plane vibrations of S and Mo atoms and out-of-plane vibrations of S atoms, respectively [40]. Multiple Raman measurements are carried out, and a typical Raman spectrum is provided in Figure 1c and a spectrum collected from other spots was provided in supplementary information in Figure S2a. Generally, to extract the Raman mode frequency value, Raman spectra were fitted with the Lorentzian function. The peak value of $E^1_{2g}$ and $A_{1g}$ mode was found to be 385.3 and 404.7 cm$^{-1}$, respectively (Figure 1c and inset of Figure S2a). Using Raman correlative analysis by considering two prominent Raman modes[41, 42], the carrier concentration of 1L-MoS$_2$ is calculated to be 1.88×10$^{12}$ cm$^{-2}$. In the MoS$_2$, the difference ($\Delta$) between two prominent Raman modes is taken to calculate the number of layers in a film [37]. Particularly, $\Delta$ value of $\leq 21$ cm$^{-1}$, corresponds to the monolayer nature of the MoS$_2$ films [43]. In this study, the $\Delta$ value is calculated to be 20 cm$^{-1}$ (inset of Figure S2a); therefore, the as-grown MoS$_2$ film is confirmed to be a monolayer. To assess the optical quality of the grown sample, PL spectroscopic measurements were carried out. A typical PL spectrum of MoS$_2$ in RT is manifested by emission from neutral excitons and trions [36-38]. In 1L-MoS$_2$, the intense A-exciton emission around 1.84 eV confirms the direct band gap nature of 1L-MoS$_2$, and with increasing the layer number, the MoS$_2$ behaves as indirect band gap material with a drastic reduction in PL emissions [44]. A typical PL spectrum is provided in Figure 1d and another PL spectrum is provided in supporting Information (Figure S2b). Generally, the PL spectrum is fitted with the Voigt function to extract the corresponding emission energy. The fitting results reveal that the PL spectrum of a typical MoS$_2$ sample (Figure S2b) consists of emissions from trions (1.80 eV), A-excitons (1.82 eV), and B-excitons (2.0 eV). In the present study, the PL emission is dominantly contributed by A-exciton at energy ~1.82 eV, therefore, the as-grown MoS$_2$ film was confirmed to be a monolayer with good optical quality. In addition, we mapped the A-exciton emission (intensity and peak energy) for PL imaging, and the Raman shifts of the characteristic in-plane ($E^1_{2g}$) and out-of-plane ($A_{1g}$) Raman modes for Raman imaging of 1L-MoS$_2$ (Figure S2d-f). The resulting maps show spatially uniform A-exciton intensity (Figure S2d) and nearly constant Raman peak positions (Figure S2e, f) across the entire scanned region, with negligible spatial variation. These data confirm that the MoS$_2$ film is continuous and homogeneous in thickness.

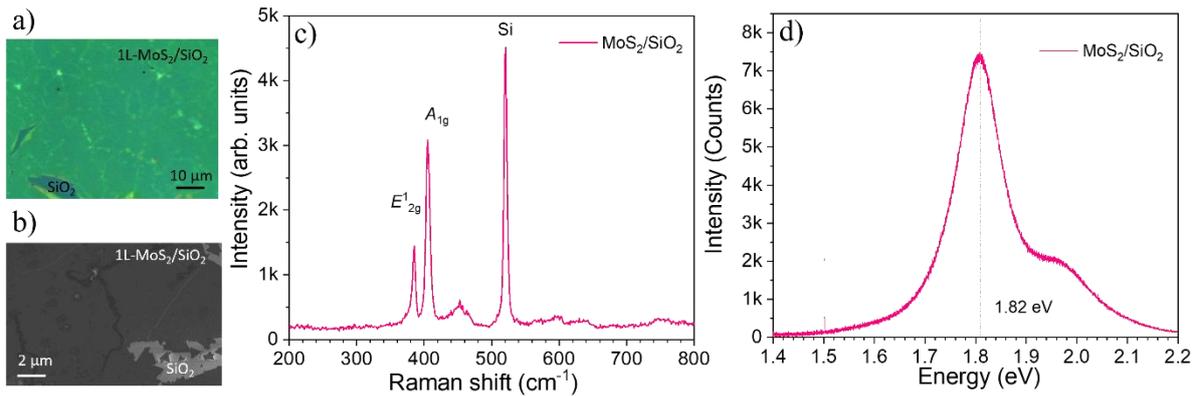

**Figure 1:** Optical characterization of CVD-grown 1L-MoS$_2$ film on SiO$_2$/Si substrates. (a) Typical optical microscopic image of large-area 1L-MoS$_2$, (b) FESEM micrograph showing the uniform and large-area 1L-MoS$_2$ film. A typical (c) Raman and (d) PL spectrum of as-grown 1L-MoS$_2$ film on SiO$_2$/Si substrate.

After the successful growth and characterization of 1L-MoS$_2$ on SiO$_2$/Si substrates, we transferred centimetre-scale 1L-MoS$_2$ films onto commercially procured low-temperature-grown GaAs substrates using the most practiced surface-energy-assisted transfer method. Before that, the commercially purchased GaAs substrates were cleaned thoroughly, and their optical characterizations were performed. Particularly, Raman and PL measurements were carried out for GaAs substrates (Figure S3). The prominent Raman modes such as TO (265 cm$^{-1}$), and LO (290 cm$^{-1}$) modes correspond to the signature peaks for GaAs [35]. Moreover, strong PL emissions with energy at 1.42 eV correspond to the good optical quality of the GaAs substrate (Figure S3). The 1L-MoS$_2$ film was transferred onto GaAs substrates using the polystyrene (PS) polymers as carriers from 1L-MoS$_2$ on SiO$_2$/Si substrate [20, 22]. The detailed film transfer methodologies are discussed in the experimental sections. A step-by-step 1L-MoS$_2$ film transfer protocol is presented as a schematic, and a few captured photographs are provided in Figure S4. The centimeter-scale 1L-MoS$_2$ film was successfully transferred from SiO$_2$/Si onto the GaAs substrate (Figure S4). The optical and FESEM micrographs containing both GaAs and 1L-MoS$_2$/GaAs heterostructures were studied to reveal the contrast. The optical image reveals the large-area, uniform, and continuous 1L-MoS$_2$ film on the GaAs substrate (Figure 3a). Similarly, to complement the optical microscopic result, a high-magnified FESEM imaging is carried out (Figure 3b). The surface morphology of the fabricated heterostructure is found to be uniform. To confirm the phase of the heterostructure, multiple micro-Raman measurements were carried out and spectra were plotted in Figure 2c and Figure S6a. Four prominent Raman modes at 265, 290, 382.5, and 404 cm$^{-1}$ were observed. Among them, Raman modes at 265 cm$^{-1}$ (TO), and 290 cm$^{-1}$ (LO) peaks correspond to the signature peaks for GaAs fingerprints, and peaks at 382.5 cm$^{-1}$ ($E^1_{2g}$), and 404 cm$^{-1}$ ($A_{1g}$) correspond to 1L-MoS$_2$ vibrational modes (Figure 2c). Using Raman correlative analysis by considering two prominent Raman modes [41, 42], the carrier concentration of 1L-MoS$_2$ is calculated to be 8.89×10$^{11}$ cm$^{-2}$. The one-order decrease in carrier concentration in constituent 1L-MoS$_2$ at heterostructures indicates the charge transfer from 1L-MoS$_2$ to GaAs. The observation of characteristic Raman peaks of both materials confirms the formation of a heterostructure.

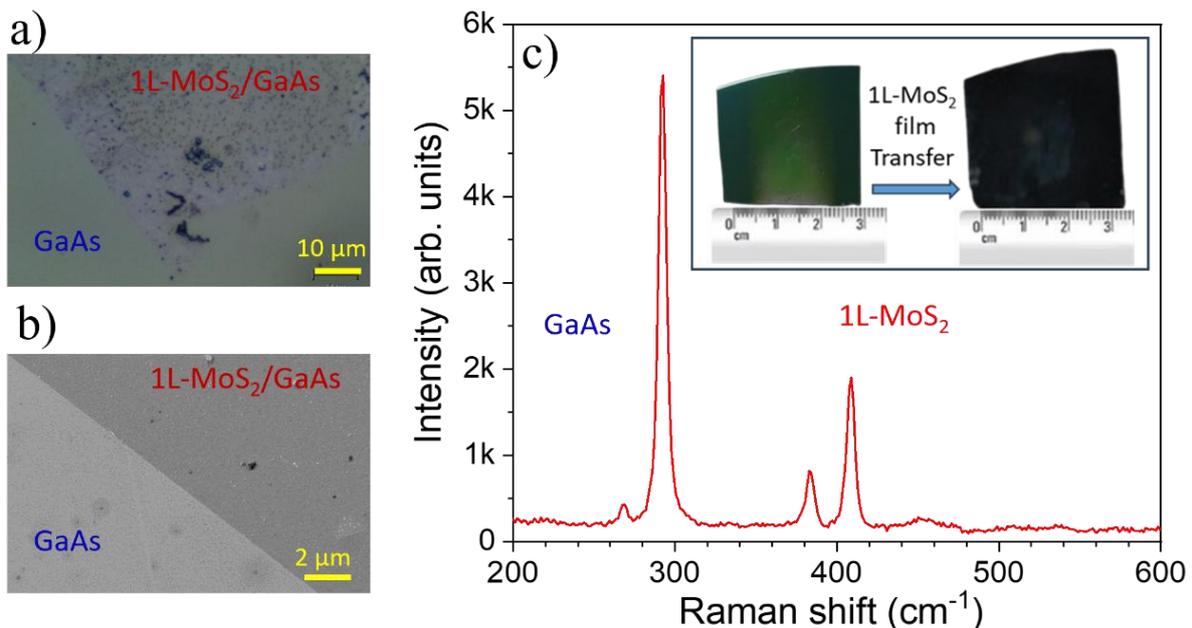

**Figure 2:** Fabrication of large-area 1L-MoS$_2$/GaAs heterostructures. (a) Optical micrograph showing the contrast between GaAs and 1L-MoS$_2$/GaAs area (b) FESEM micrograph shows the morphology of the bare GaAs and heterostructures, (c) Raman spectrum of 1L-MoS$_2$/GaAs heterostructure, showing signature peaks of both 1L-MoS$_2$ and GaAs. The inset shows the photographs of a 1L-MoS$_2$/SiO$_2$ and 1L-MoS$_2$ transferred onto GaAs substrate, to fabricate 1L-MoS$_2$/GaAs heterostructure.

To understand the emission properties of the fabricated heterostructure, the micro-PL measurements were carried out using 532 nm continuous wave (CW) laser excitation at room temperature. To avoid the photodoping effect, we used a significantly low laser power (≤100 µW) for the PL measurements. For a fair comparison, the nearest spots with and without 1L-MoS$_2$ were selected for PL measurements. The absence of 1L-MoS$_2$ on GaAs was confirmed by optical microscopic and Raman spectroscopic measurements. The multiple PL spectra were collected from the heterostructure (1L-MoS$_2$/GaAs) and GaAs (spots without 1L-MoS$_2$) and shown in Figure 3a and Figure S6. The PL spectra of 1L-MoS$_2$ and its heterostructures were fitted with Gaussian functions, yielding a consistent linewidth of 27 meV (full-width at half-maximum, FWHM) for both systems (Figure S7). The unchanged linewidth in the heterostructure strongly suggests that radiative recombination dominates the excitonic decay process. This observation implies a lack of significant non-radiative pathways, such as carrier trapping at interfacial states or defect sites, which would otherwise broaden the PL linewidth by introducing additional recombination channels. The preserved linewidth further highlights the high-quality interface in the heterostructure, with minimal defect-mediated scattering. However, as a commercially procured GaAs substrate are grown at low temperature point defects may exist, which may facilitate efficient interfacial charge transfer and the monolayer MoS$_2$ can inject or transfer excitations into a GaAs host that quickly traps and neutralizes carriers, leading to transient charge separation currents that emit THz radiation [45].

Heterostructures show about two times enhancement of PL emission intensity at 1.42 eV, compared to the bare GaAs substrate (Figure 3a). Generally, the enhancement factor is calculated by the integrated intensity of both acceptor and donor material to the intensity of bare acceptor material [46]. Using the integrated intensity of the PL spectra of heterostructures and GaAs (Figure S7), the enhancement factor is calculated and found to be ~2.38. Interestingly, a low-intense peak at an emission energy of 1.84 eV is also observed (insets of Figure 3a). A blue shift of ~20 meV in PL emission corresponds to the release of compressive strain introduced during the CVD growth of 1L-MoS$_2$. The substantial reduction of PL intensity at 1.84 eV and enhancement of the PL intensity at 1.42 eV reveal the dominant Dexter-type exciton transfer at type-I band alignment. The attribution of Dexter-type exciton transfer in the 1L-MoS$_2$/GaAs heterostructure is supported by two critical factors: (1) spectral overlap, and (2) atomic-scale proximity. The excitonic emission energy of 1L-MoS$_2$ (~1.84 eV) aligns with the absorption spectrum of GaAs (direct bandgap ~1.42 eV) (Fig. 3b). Phonon-assisted processes or defect-mediated transitions likely bridge the small energy mismatch (~0.42 eV), enabling resonant coupling. The proximity of two intralayer excitons can be separated by the van der Waal gap between two semiconductors. The total thickness of the heterostructure (1L-MoS$_2$ and interfacial van der Waals gap) was measured via atomic force microscopy (AFM) as 0.85 nm. (Fig.3c). Subtracting the known thickness of transferred 1L-MoS$_2$ (~0.7-1.1 nm, consistent with the literature [47, 48], the interfacial gap is estimated to be <0.15 nm.

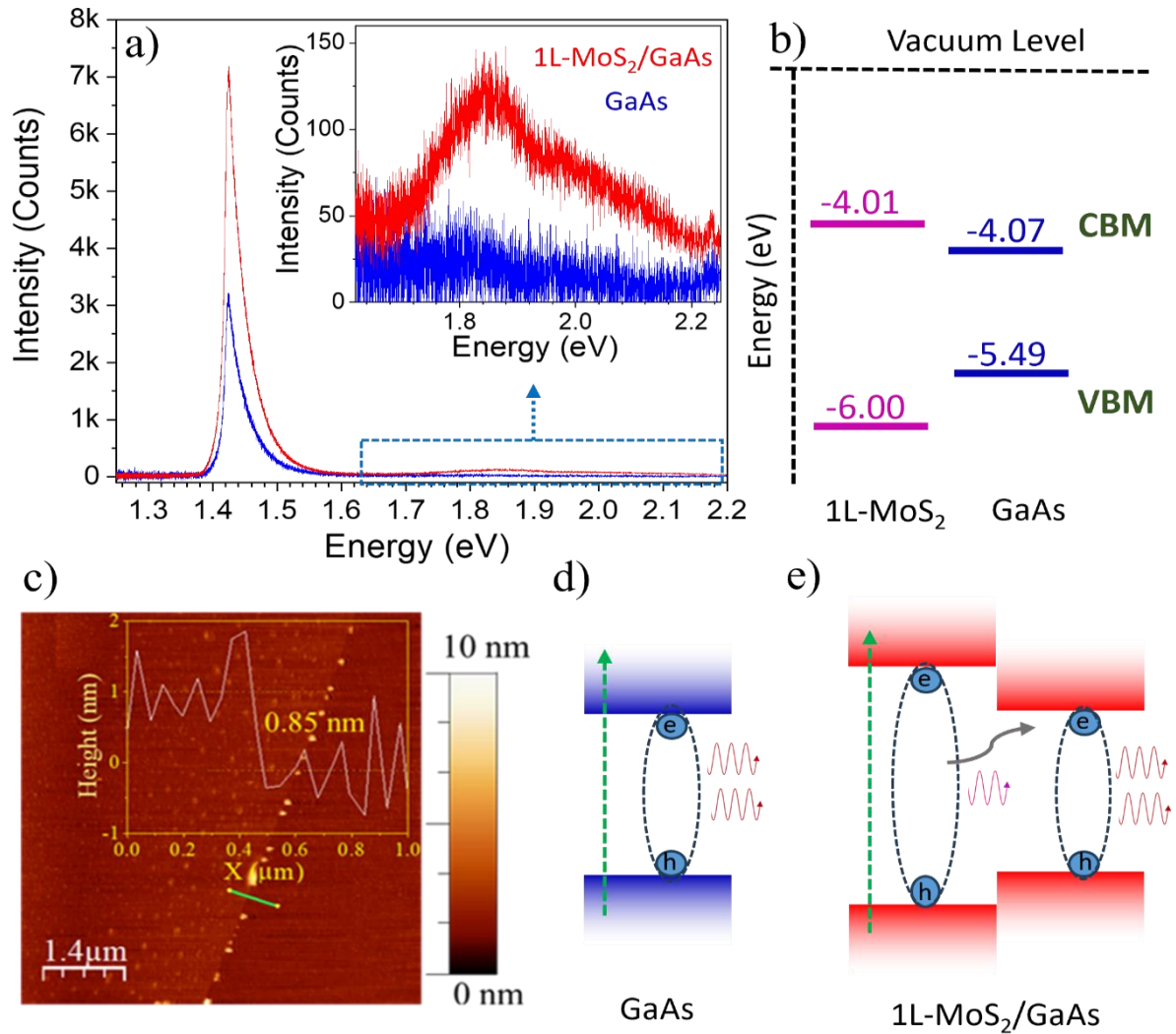

**Figure 3:** Photoluminescence studies of 1L-MoS$_2$/GaAs heterostructures. (a) Typical PL spectra were collected from areas containing and without 1L-MoS$_2$ in the energy range covering 1.2-2.2 eV. Inset showing PL emissions at 1.84 eV with low emission intensity. (b) Type-I band alignment cartoon showing valence band maxima (VBM) and conduction band minima (CBM) of both GaAs and 1L-MoS$_2$ semiconductors. (c) Atomic-force microscopy image of transferred 1L-MoS2 on GaAs substrate. The inset shows the total thickness of the film on GaAs, which helps to calculate the van der Waals gap. Illustration of photoluminescence emissions in (d) GaAs and (e) 1L-MoS$_2$/GaAs heterostructure, where the generated excitons of 1L-MoS2 at an energy of 1.84 eV are transferred to GaAs, which emit at 1.42 eV.

The photophysical processes in the 1L-MoS$_2$/GaAs heterostructure were probed using a green laser (2.33 eV), which exceeds the electronic bandgap of monolayer MoS$_2$ (~1.99 eV). This bandgap comprises the neutral A-exciton energy at ~1.84 eV and a binding energy of ~150 meV, characteristic of strongly bound excitons in 2D transition metal dichalcogenides [49]. Due to its atomic-scale thickness, monolayer MoS$_2$ exhibits reduced optical absorption (~10 % of incident visible light compared to bulk counterparts) [48], yet the high photon energy of the green laser generates a significant density of photocarriers (electron-hole pairs) within the 1L-MoS$_2$ layer. These photocarriers rapidly form excitons, which subsequently transfer to the GaAs substrate via a Dexter-type mechanism (Fig. 3d,e). This process involves the simultaneous two-particle exchange of an electron and a hole across the heterointerface [50]. The energy transfer rate ($k$) of Dexter-type is given by $k = KJ\exp(-2d/L)$, where J represents the normalized spectral overlap integral, $K$ is an experimental factor, $d$ is the gap between donor and acceptor, and $L$ is the sum of the vdWs radius of the donor and acceptor. The proximity between MoS$_2$ and GaAs atoms at the interface directly affects orbital overlap. Smaller interlayer distances enhance wavefunction overlap, increasing the exchange integral. Particularly, van der Waals interactions typically govern this spacing in heterostructures. From the theoretical calculation, the work function and its corresponding band maxima and minima for both 1L-MoS$_2$ and GaAs are considered (Figure 3b). The work function of 1L-MoS$_2$ and GaAs is mentioned to be 4.01 and 4.07 eV, respectively. The built-in potential at the interface is calculated to be 60 meV, which is compensated by a phonon-assisted process. The phonons generated at the interface are due to excess energy, which is the energy difference in the exciton emission energy of MoS$_2$ and the bandgap of GaAs. In the existing literature, the heterostructures using bulk MoS$_2$ and GaAs reveal the transfer of electrons from GaAs to bulk MoS$_2$ [51]. However, our result solidifies a strong understanding of the transfer of photoexcited electrons and holes from 1L-MoS$_2$ to GaAs lattice through a Dexter-type exciton transfer mechanism. Overall, the 1L-MoS$_2$ acts as an efficient visible light-absorbing material to enhance the emission efficiency for GaAs-based devices.

To investigate the occurring multiple linear and non-linear optical phenomena at the heterostructure due to an exciting sample with an ultrafast femtosecond laser beam, terahertz time-domain (THz-TDS) measurements are performed [52]. The impact of generated photocarriers at the heterointerface (1L-MoS$_2$/GaAs) is studied by THz-TDS measurements. The measurement setup of THz generation and detection is provided in the experimental section in detail. The generated THz pulses pass through the bare GaAs and 1L-MoS$_2$/GaAs heterostructure, and their time domain signal is plotted in Figure 4a. A few THz-TDS spectra are also provided in Figure S8. The collected time-domain THz pulses are converted to the frequency-domain through Fast Fourier Transformations (FFT), and further, the frequency-dependent optoelectronic parameters such as refractive index, absorption coefficient, and optical conductivity are plotted. As film thickness ($d$) is less than the thickness of the substrate ($d_{sub} > d$) and the wavelength of the probe, $\lambda_{inc}$ is greater than $d$; therefore, we used Tinkham approximation to extract all optical parameters; we used the Tinkham approximation to extract all optical parameters [53, 54]. The refractive index ($n$) in the range of 0.1-2 THz is plotted in Figure 4b. In bare GaAs, the $n$ value varies in the range of 3.0-3.5. However, the value is reduced to 1.6-2.5 at heterostructure (1L-MoS$_2$/GaAs). The decrease in the $n$ value for the heterostructure is attributed to the variation of free charge carriers due to exciton transfer at the interface. The charge exchange can also manipulate the dielectric constant of the heterostructure, which is different from the bare GaAs. Similarly, we have also extracted the absorption coefficient ($\alpha$) and plotted it in Figure 4c. The increase in the $\alpha$ value with increasing frequency for GaAs is due to the contribution of both charge carriers and phonons [55]. The abrupt change in $\alpha$ value around ~1 THz likely arises from the interplay of different THz generation and propagation mechanisms in our heterostructure. One plausible explanation is the crossover between two dominant THz emission processes at the MoS$_2$/GaAs interface: (i) ultrafast transient photocurrents and (ii) nonlinear polarization effects. Each mechanism has a characteristic frequency response, and their superposition can lead to a non-monotonic spectrum [56]. Indeed, prior studies on THz emission from layered semiconductors have reported that at lower THz frequencies the emission is often dominated by photocurrent surge, whereas at higher frequencies optical rectification becomes significant [56]. In our MoS$_2$/GaAs heterostructure, a similar situation may occur near 1 THz. At frequencies below ~1 THz, the THz waves are likely dominated by the GaAs photocarrier surge current, whereas above ~1 THz, the interfacial polarization or excitonic nonlinearities in the monolayer start to contribute altering the slope of the absorption or emission spectrum. At 2 THz, the $\alpha$ value is found to be 35 and 5 cm$^{-1}$ for bare GaAs and 1L-MoS$_2$/GaAs, respectively. The decrease in the $\alpha$ value delineates the increase in the optical transmittance for the heterostructure compared to bare GaAs, which is further attributed to the strong interference generated at the interface. The 78 % decrease in the $\alpha$ value in the heterostructure, compared to the bare GaAs substrate, may originate from optical loss. In particular, the optical loss also covers both absorption loss and scattering loss that occur at the nanoscale heterointerface. Further, the real part of optical conductivity ($\sigma$) can be extracted by using frequency-dependent refractive index ($n$) and transmittance, T($\omega$), using the relations [54]

$$\sigma(\omega) = \frac{1+n}{Z_0} \left(\frac{1}{T(\omega)} - 1\right) \text{ where } T(\omega) = \frac{E(1L-MoS2/GaAs)}{E(GaAs)}$$

At 2 THz, the σ value is calculated to be ~6 and ~9 S/m for bare GaAs and 1L-MoS$_2$/GaAs, respectively. The increase in optical conductivity in the heterostructure is attributed to the participation of interfacial charge carriers along with quasi-particles of 1L-MoS$_2$. The generated photoexcited electrons in 1L-MoS$_2$ are transferred to GaAs to enhance the overall optical conductivity of the heterostructure. The out-of-plane current density is also increased in similar MoS$_2$/WS$_2$ heterostructures [17].

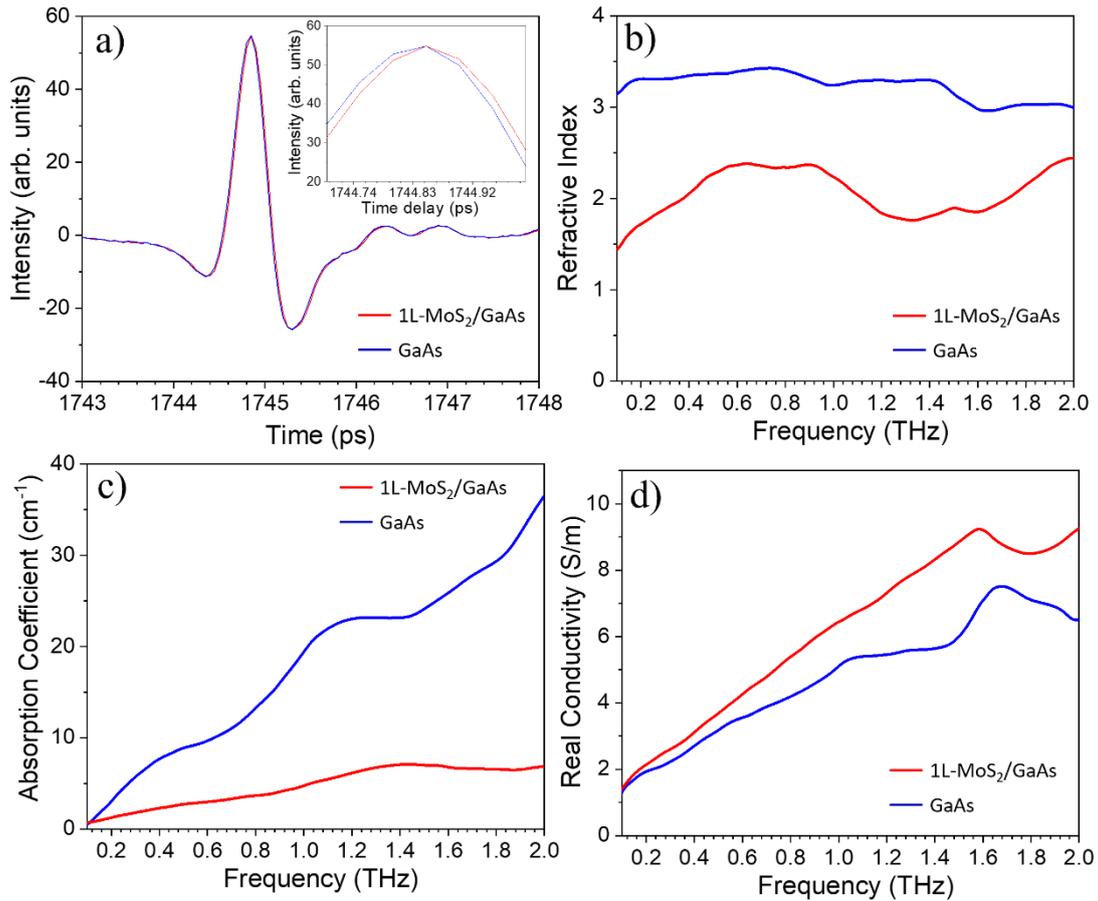

**Figure 4:** Optoelectronic parameters extracted from the THz-TDS measurements in the range of 0.1-2 THz range, for both GaAs substrate and heterostructures (1L-MoS$_2$/GaAs). (a) THz-TDS signal in transmission configurations. The inset shows the magnified THz-TDS signal with a peak around ~1744.83 ps. (b) refractive index, (c) absorption coefficient, and (d) real optical conductivity of both bare GaAs and 1L-MoS$_2$/GaAs heterostructures.

THz generation in heterostructures is performed to investigate how interfacial effects and ultrafast response play a critical role at these interfaces, and the detailed THz measurement setup and protocols are outlined in the experimental section. Moreover, a schematic of the setup for THz generation for both bare GaAs substrate and the heterostructure (1L-MoS$_2$/GaAs) is depicted in Figure 5a, while the generated output THz power with increasing laser power is presented in Figure 5b. The signal efficiency is calculated based on our previous reports [57, 58].

The output power and signal efficiency were found to be increased (~15%) for the heterostructure than for the bare GaAs substrate (Figure 5c).

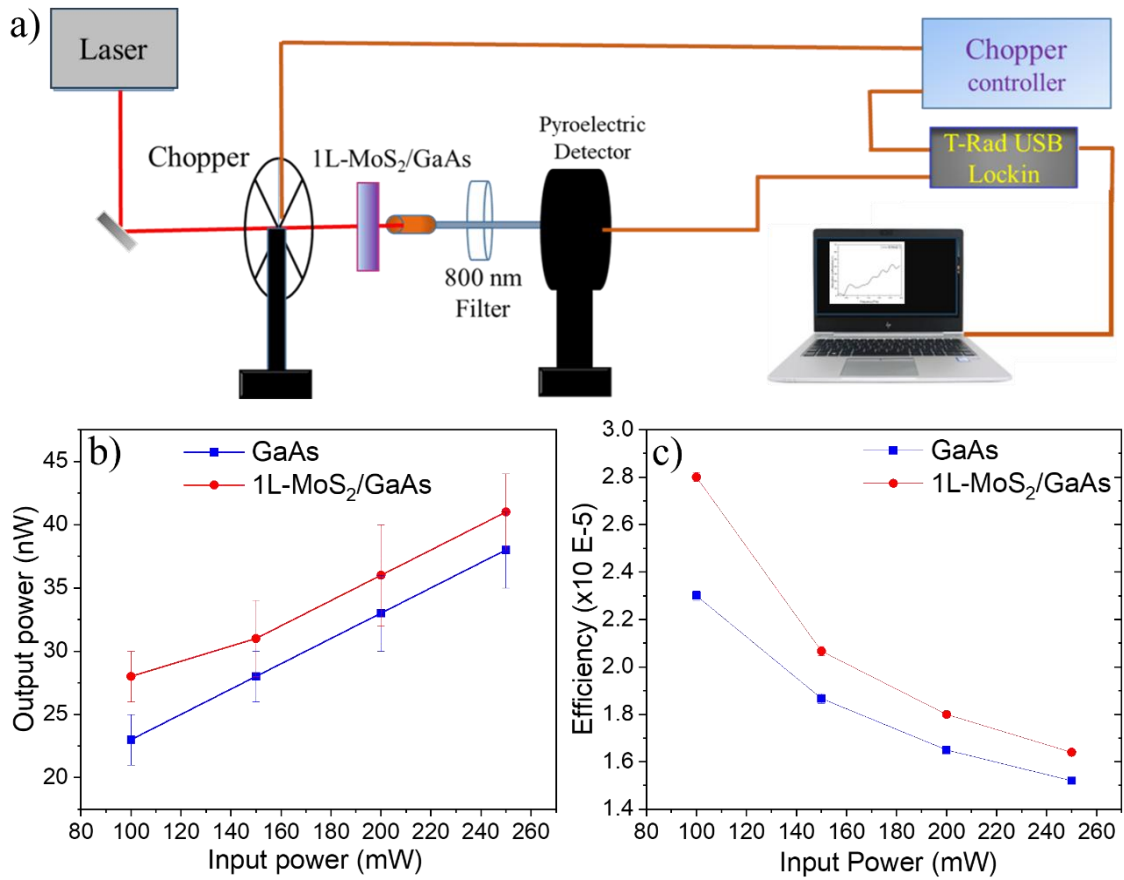

**Figure 5**: THz generation measurements. (a) The schematic of the optical setup was designed for THz generation. (b) Incident laser power versus output THz power plot and corresponding (c) calculated efficiency for both GaAs and 1L-MoS$_2$/GaAs heterostructure.

The increase in THz generation is due to a synergistic nonlinear response in the heterostructure. Therefore, it is important to understand the non-linear optical response of their constituent materials and their constructive interference phenomena. 1L-MoS$_2$ shows strong in-plane nonlinear optical properties due to its non-centrosymmetric hexagonal crystal structure and contribution from second-order nonlinear susceptibility arising from broken inversion symmetry [59]. Similarly, GaAs is a non-centrosymmetric cubic structure with inherent second-order non-linear susceptibility ($\chi^{(2)}$) properties. It also exhibits optical rectification, where ultrafast photoexcitation generates a directional surge photocurrent, producing broadband THz waves [60]. In a heterostructure, the generated output THz power is mainly contributed by two factors: (1) Phase-coherent addition of $\chi^{(2)}$, and (2) Overall enhancement in the electric field of the laser. In phase-coherent addition of $\chi^{(2)}$, the $\chi^{(2)}$ tensors of MoS$_2$ and GaAs may align spatially and interfere constructively, which results in the enhancement of the overall $\chi^{(2)}$ value.

Also, the redistribution of the laser electric field also increases the effective electric field in GaAs, 1L-MoS$_2$, and their interfaces. The generated THz power, $P_{THz} \propto \chi^{(2)} E^2$, where $E$ is the electric field of laser [61]. In a heterostructure, the sum of $P_{THz}$ contributed from both 1L-MoS$_2$ and GaAs increases the overall output THz power than bare GaAs.

Our results were benchmarked against previously reported studies. Zhang *et al.*, reported a graphene-silicon photoconductive switch that achieved an 80× enhancement in THz amplitude over a conventional Si emitter [45]. In their design, graphene acted as a "hot-carrier fast lane," rapidly extracting photoexcited carriers from silicon to overcome the trade-off between carrier lifetime and mobility [45]. The biased photoconductive approach demonstrates the extreme case of a 2D/3D interface improving THz output (by orders of magnitude). In comparison, our approach uses an unbiased surface emitter (no electrical bias applied) and relies on exciton transfer from monolayer MoS$_2$ to GaAs. Similarly, another study explored coating GaAs with ultrathin conductive films or nanostructures to boost THz emission. For instance, Sahoo *et al.*, demonstrated a ~2.5× increase in THz peak field by depositing a rough indium tin oxide (ITO) film on a semi-insulating GaAs substrate [56]. The enhancement was attributed to plasmonic local field concentration at the ITO/GaAs interface and constructive interference of THz waves emitted from the interface [56]. However, unlike a continuous conductive oxide film, the monolayer semiconductor provides a unique excitonic contribution: it can absorb photons and generate excitons that transfer energy into the GaAs, potentially yielding additional ultrafast carriers. Moreover, MoS$_2$ is atomically thin and minimally perturbative to the THz propagation, which is an advantage over thicker coatings that might introduce attenuation. We also compare our results to studies on van der Waals heterostructures used in THz emission experiments. For example, THz surface emission has been observed in WS$_2$/MoS$_2$ bilayers, where an interlayer charge transfer current at the interface contributes to THz generation [62]. Those works underscore that interface-generated currents (even without an applied bias) can be a significant THz source in 2D/2D systems. In our 2D/3D system, the principle is similar: the MoS$_2$/GaAs junction forms a type-I band alignment, facilitating charge or energy transfer across the interface. We emphasize that a key advantage of our approach is the combination of a strong light absorber (1L-MoS$_2$) with an efficient THz emitter (GaAs). As mentioned in the introduction, 1L-MoS$_2$ has a high exciton binding energy and absorbs in visible wavelengths, whereas GaAs (especially GaAs) excels at converting absorbed photons (or transferred excitons) into ultrafast photocurrent transients. By comparing to the literature, we highlight that for the first time, to the best of our knowledge, we have demonstrated enhanced THz emission via exciton transfer in a TMD/GaAs heterostructure. The observation extends the paradigm of 2D/3D hybrids for THz generation. Further, the understanding of carrier dynamics of constituent materials and interfacial contributions can be explored in the future.

## 3. Conclusions

We have successfully grown large-area 1L-MoS$_2$ on SiO$_2$/Si substrates using the CVD method. Further, the monolayer films were transferred onto a GaAs substrate by the surface-energy-assisted transfer method to fabricate a 1L-MoS$_2$/GaAs van der Waals heterostructure. The heterostructure shows 15% higher THz generation compared to bare GaAs, which is attributed to the enhancement of the optical rectification process in GaAs using 1L-MoS$_2$ film. The enhancement in the generated THz signal is understood from the phase-coherent addition of χ(2), and the overall enhancement in the electric field of the laser. The optical conductivity is increased due to the generation of additional free-charge carriers. Further, it reveals that the heterojunction is type-I band bending with Dexter-type exciton transfer and participation of phonons. The exciton transfer is well supported by photoluminescence spectroscopy. Transferred excitons dissociate in GaAs, enhancing radiative recombination efficiency. The novel mixed-dimensional van der Waals heterostructure has the potential to enable many exciting opportunities to design compact and efficient THz nanodevices with on-demand tunable optoelectronic properties.

## 4. Experimental section

**Fabrication of 2D/3D Heterostructures (1L-MoS$_2$/GaAs)**

We synthesized the centimeter-scale 1L-MoS$_2$ film on SiO$_2$/Si substrates using the chemical vapor deposition (CVD) method. To form the MoS$_2$ phase, we loaded 15 mg of MoO$_3$ (99.9%, Sigma-Aldrich) and 60 mg of S (99.9%, Sigma-Aldrich) powder into an alumina boat, placing them in two extreme zones of a three-zone furnace. MoS$_2$ phase formation was achieved by flowing 50 sccm of ultra-high purity (UHP) Ar gas for 3 h. Further details regarding the synthesis steps can be found elsewhere [36]. The GaAs substrates used in our study were commercially procured low-temperature-grown GaAs wafers. Typically, GaAs was epitaxially grown at approximately 200-300 °C to incorporate a high density of defects [63]. We employed a well-established 2D film transfer process, specifically surface-energy-assisted transfer techniques, to transfer the 1L-MoS$_2$ film onto GaAs substrates. Detailed protocols for the film transfer can be found elsewhere[64].

**Optical and morphological characterization**

Raman and photoluminescence (PL) measurements were conducted by a micro-Raman spectrometer (inVia, Renishaw, UK). The spectrometer is equipped with a 532 nm laser source for exciting the 1L-$MoS_2$ and heterostructures. The scattered light is then dispersed using gratings with 2400 lines/mm and 1800 lines/mm for Raman and PL measurements, respectively. Subsequently, the dispersed light is captured by a charge-coupled device detector operating at room temperature. The morphology of the heterostructure was studied using a FESEM (Supra 55, Zeiss, Germany).

**THz-TDS measurements**

The THz TDS measurements were performed in transmission mode, utilizing the TERAFLASH THz system. The femtosecond laser employed (TOPTICA FemtoFErb FD6.5) operated at a wavelength of 1560 nm, with a repetition rate of 100 MHz and a pulse duration of 50 fs. The output power was approximately 72 mW, and the diameter of the THz beam was approximately 2 mm. The femtosecond laser pulse is incident on the sample normally. To avoid humidity intervention, all TDS measurements were carried out in an in-house designed dry air-purged system. Also, the precision of all measurements was calibrated for the Au-coated mirror. The THz pulses of 0.3 ps at 100 MHz repetition rate were used for all measurements. The optoelectronics parameters, such as refractive index, absorption coefficient, and optical conductivity, were calculated using frequency domain data.

**THz Radiation Generation and Detection Setup**

The power of the generated THz radiation was quantified using a custom-built THz generation apparatus. THz radiation originating from 1L-$MoS_2$/GaAs heterostructure was produced utilizing a configuration linked to a Ti: sapphire laser amplifier (COHERENT LIBRA), operating with the following parameters: pulses at 800 nm wavelength lasting 50 fs, at a repetition rate of 1 kHz. The incident laser power was controlled through a half-wave plate attenuator. The resulting radiation comprised a mixture of THz radiation and unchanged 800 nm wavelength, which were subsequently separated via Teflon and Si filters. Measurement of the generated THz power was conducted utilizing a pyro-detector (Genetic Made) at ambient. The detector output was directed to a lock-in amplifier, and the power of the generated signal was determined using dedicated software. For detection and measurement of the THz radiation power, a pyroelectric detector, integrated with a built-in lock-in amplifier operating at a frequency of 25 Hz, was chosen due to the sluggish response of the thermoelectric sensor crystal. This configuration not only improved the signal-to-noise ratio but also mitigated the impact of surface damage on thin films. The incident power of the 800 nm wavelength laser was systematically varied within the range of 100 to 250 mW, with increments of 50 mW. The entire experimental procedure was replicated for bare GaAs substrates as well. Multiple measurements were carried out to obtain better statistical accuracy.

**Supporting Information**

Supporting Information is available from the Online or from the author.

**Acknowledgment**

The authors gratefully acknowledge the financial support provided through the Post-Doctoral Research Fellowship of the Institute of Eminence, University of Hyderabad. We thank R. N. Vamsi Krishna (University of Hyderabad) for assistance with the preliminary THz generation measurements. We also thank Dr. S. Amrithapandian (IGCAR) for performing the FESEM measurements, and the Director, IGCAR, for providing access to the facilities used for monolayer film growth and transfer.

**Conflict of Interest**

The authors declare no conflict of interest.

**Data Availability Statement**

The data that support the findings of this study are available in the Supporting Information of this article.